\documentclass{article}
\usepackage{natbib}

\begin{document}

\bibliographystyle{crs}
\title{A Simple Model of the Evolution of Simple Models of Evolution}
\author{Cosma Rohilla Shalizi\thanks{Corresponding Author} \\ {\small {\it Physics Department, University of
Wisconsin, Madison, WI 53706}}\\
{\small {\it and the Santa Fe Institute, 1399 Hyde Park Road, Santa Fe, NM 87501}}\\
{\small {\it shalizi@santafe.edu}}
\and William A. Tozier \\ {\small {\it William Tozier Consulting, PO Box 8002, Ann Arbor, MI 48107}}\\
{\small {\it and The Santa Fe Institute}}\\
{\small {\it bill@williamtozier.com}}}
\date{12 October 1999}

\maketitle

\begin{abstract}
In the spirit of the many recent simple models of evolution inspired by
statistical physics, we put forward a simple model of the evolution of such
models.  Like its objects of study, it is (one supposes) in principle testable
and capable of making predictions, and gives qualitative insights into a
hitherto mysterious process.
\end{abstract}

\vspace{10mm}

Even the most casual perusal of {\it Physical Review Letters}, {\it Physical
Review E}, {\it Journal of Statistical Physics} or the Los Alamos e-print
archive for nonlinear systems could not fail to reveal that the last decade,
indeed, the last few years, have seen a remarkable explosion of simple models
of biological evolution formulated by physicists.  These draw their inspiration
not from the founding works on the mathematical modeling of evolutionary
processes (\cite{Fisher-genetical}, \cite{Haldane-causes} and
\cite{Sewall-Wright-papers}), nor from the extensive development and refinement
of this theory by six subsequent decades of active research ({\it e.g.},
\cite{Hamilton-narrow-roads}, \cite{Maynard-Smith-games,Maynard-Smith-evo-gen},
\cite{Hofbauer-Sigmund}), nor the highly abstract \cite{Holland-adaptation},
nor even the less orthodox biologists ({\it e.g.}, \cite{Kauffman-origins}) but
from models well-established in statistical physics --- sandpiles,
reaction-diffusion systems, phase transitions, \&c., \&c.  Cynics have said
that mathematical physics is that which hasn't enough rigor and generality to
be math, but not enough contact with reality to be physics; fortunately we are
not cynics, or we would not be able to resist making the analogous {\it bon
mot} about this new wave of physico-biology \cite[]{Lotka}.

The question presents itself: why are we being deluged with such models?  In
the spirit of the field, we present a simple evolutionary model of this
process.

\begin{enumerate}
\item A physicist runs across or concocts from whole cloth a mathematical model
which is simple, neat, and contains a great many variables of the same sort.
\item The physicists has heard of \cite{Darwin-origin}, and may even have read
\cite{Dawkins-watchmaker} or some essays by Gould, but wouldn't know
\cite{Fisher-genetical}, \cite{Haldane-causes} and \cite{Sewall-Wright-papers}
from the Three Magi, and doesn't dream that such a subject as mathematical
evolutionary biology exists.
\item The physicist is aware that lots of other physicists are interested in
annexing biology as a province of statistical physics.
\item The physicist interprets his multitude of variables as species or (if
slightly more sophisticated) as {\it genotypes}, and proclaims that he has
found ``Darwin's Equations'' ({\it cf.}
\cite{Bak-Flyvberg-Sneppen-new-scientist}), or, more modestly, has made an
important step towards eventually finding those equations.
\item His paper is submitted for review to other physicists, who are just as
ignorant of biology as he, but see that it's about equivalent to the other
papers on evolution by physicists.  They publish it.
\item The paper is read by other physicists, because at least it's not another
derivation of specific heats on some convoluted lattice under a Hamiltonian
named for some Central European worthy now otherwise totally forgotten.
Said physicists think this is cutting-edge evolutionary theory.
\item Some of those physicists will know or discover simple, neat models with
lots of variables of the same type.
\end{enumerate}

A number of observations seem called for.

First, an analogous process in another field of cogno-intellectual ecology
has recently been postulated and experimentally documented by a respected
statistical physicist (\cite{Sokal-experiments,Sokal-transgressing}).  This
can only lend support to our model.

Second --- and we confess this is a flaw from the aesthetic point of view ---
our model is not completely detached from the existing literature on the
evolution of ideas.  While not strictly a memetic theory in the sense of
\cite{Dawkins-selfish} and \cite{Lynch-contagion}, it is very close in spirit
to the ``epidemiology of ideas'' proposed in \cite{Sperber-explaining}.  We do
not assume any very high degree of similarity between the simple models of
evolution, {\it i.e.}, they are not reliable replicators.  Quite the contrary,
our model predicts that, to within an order of magnitude, there will be as many
distinct models as there are modelers (allowing for collaborations and the
proposal of multiple models).  Since the models acquire relevance (in the sense
of \cite{Sperber-explaining,Sperber-Wilson}) through distinctiveness and
novelty, it could hardly be otherwise.

Third, our model predicts that simple statistical-physical models of evolution
will continue to proliferate until either (a) all the available models are
exhausted, or (b) they become as common and as boring as any other subject in
the statistical physics literature, or (c) physicists learn some actual
biology.  We are not entirely confident that the third limiting factor will
become operational before the others.

Finally, the astute reader of this note will also see that we have not
ourselves taken a statistical mechanics approach to modeling the dissemination
and diversification of physicists' evolutionary models, but have rather left
this as an exercise for subsequent modelers of models of models, though we
suspect a multiplicative noise process would be both appropriate and apt.  We
are strengthened in this suspicion by a recent investigation \cite{Redner} into
the distribution of citations of papers, independent of their subject matter,
found that they conform to Zipf's law with an exponent of $\approx -0.5$, and
the classical explanation of such phenomena, first provided by
\cite{Simon-skew}, is, precisely, multiplicative noisy growth.  Thus, were the
field of modeling modelers to come into existence, it would be rife with
potential for publication.

\subsection*{Acknowledgments}
The authors thank Marc Abrahams and Nigel Snoad for helpful discussion, and
Prof. Per Bak for providing a continual source of inspiration.  CRS thanks
Prof. David Griffeath and the undergraduate students at Madison for providing
financial support, and Prof. Yuri Klimontovich, whose book
(\cite{Klimontovich}) first alerted him to the possibilities of simple models
of evolution by physicists.  WAT has already thanked an undisclosed set of
people of analytically determined size, and expects eventually to be thanked by
others in kind due simply to expansionary propagation of that original
thankfulness (described in a subsequent paper, now in preparation).

\bibliography{locusts}

\end{document}